\documentclass[%
 aip,
 amsmath,amssymb,
 reprint,%
]{revtex4-1}

\usepackage[T1]{fontenc}
\usepackage{color}
\usepackage{array}
\usepackage{float}
\usepackage{multirow}
\usepackage{amsmath,amssymb}
\usepackage{graphicx}
\usepackage{color}

\usepackage{dcolumn}
\usepackage{bm}

\usepackage{mathptmx}

\usepackage{mathtools}

\usepackage{graphicx}
\usepackage[export]{adjustbox}
\usepackage{array}

\renewcommand{\vec}[1]{\boldsymbol{#1}}
\newcommand{\<}{\langle}
\renewcommand{\>}{\rangle}

\newcommand{\fone}{f_1}
\newcommand{\fthree}{f_3}

\begin{document}

\title{The rise and fall of branching: a slowing down mechanism  in relaxing  wormlike micellar networks}

\author{Marco Baiesi}
\email{marco.baiesi@unipd.it}
\affiliation{Dipartimento di Fisica e Astronomia, Universit\`a di Padova, via Marzolo 8, I-35131 Padova, Italy}
\affiliation{INFN, Sezione di Padova, via Marzolo 8, I-35131 Padova, Italy}

\author{Stefano Iubini}
\affiliation{Consiglio Nazionale delle Ricerche, Istituto dei Sistemi Complessi, via Madonna del Piano 10, I-50019 Sesto Fiorentino, Italy}
\affiliation{INFN, Sezione di Firenze, 
  via G. Sansone 1, I-50019 Sesto Fiorentino, Italy}

\author{Enzo Orlandini}
\affiliation{Dipartimento di Fisica e Astronomia, Universit\`a di Padova, via Marzolo 8, I-35131 Padova, Italy}
\affiliation{INFN, Sezione di Padova, via Marzolo 8, I-35131 Padova, Italy}

\begin{abstract}
A mean-field kinetic model suggests that the relaxation dynamics of wormlike micellar networks is a long and complex process due to the problem of reducing the number of free end-caps (or dangling ends) while also reaching an equilibrium level of branching after an earlier overgrowth. The model is validated against mesoscopic molecular dynamics simulations and is based on kinetic equations accounting for scission and synthesis processes of blobs of surfactants. A long relaxation time scale is reached both with thermal quenches and small perturbations of the system. The scaling of this relaxation time is exponential with the free energy of an end cap and with the branching free energy. We argue that the subtle end-recombination dynamics might yield effects that are difficult to detect in rheology experiments, with possible underestimates of the typical time scales of viscoelastic fluids.
\end{abstract}

\maketitle

\section{Introduction}
Wormlike micelles are elongated soft structures that form in solution by
thermodynamic self-assembly of its elementary constituents, amphiphilic molecules~\cite{wei06}.
This process can give rise to linear ``living'' polymers, eventually merging into
networks of branched fibres that can grow, break,
and rejoin. The process depends on temperature, surfactant concentration,
and on externally imposed stresses as in rheological and micro-rheological experiments~\cite{cat90,wei06,lar99,jeon13,gom15,tas16,ber18,jai21}.

Experiments and molecular dynamics simulations~\cite{pad05,pad08,hug11,dha15,wan17,man18}
suggest that the rich viscoelastic properties of these systems are the result of a
multiscale dynamical process: linear growth and shrinking of single fibres, occurring at short time scales, compete continuously with rewiring and branching events, whose frequency may
depend on non local spatial rearrangements of the formed network
and whose typical time scale can be of the order of seconds~\cite{gom15,ber18}.
However, also much longer time scales are possible~\cite{in07_book_ch}.

On the theoretical side, mean-field theories, reptation dynamics
and microscopic constitutive models have been
successfully developed~\cite{cates1987reptation,cat06,tur90,gra92,dry92,tur93,may97,cat06}
to rationalize the linear viscoelastic properties of living polymers.
Normally these studies focused on the distribution of polymers' length.
The role of branching in the dynamics and viscoelasticity of wormlike micellar network was also studied~\cite{dry92,may97} but has remained not fully understood~\cite{in10,in07_book_ch}.
Steep thermodynamic costs of about $30 k_B T$ were recently reported~\cite{vog17,jia18,zou19} for the scission of wormlike structures. The actual rates for scission and recombination are not provided by these equilibrium experiments, but one can expect a very low scission rate for such large free energy differences.

To complement our understanding on the equilibration process of living polymers, from microseconds up to hours,
we propose a simple kinetic model that describes the self assembly dynamics of networks
of wormlike micelles in terms of an aggregation-fragmentation process. We focus on the fractions of micelles composing different motifs of the network (end-caps, branches, strands) rather than on the commonly studied distribution of polymers' length~\cite{cates1987reptation,dry92,cat06}.
Indeed, the model is coarse grained and its basic unit is a blob of surfactants of the order of a globular micelle.
With this minimal representation we can faithfully describe the relevant elementary mechanisms as linear growing, scission, fusion and branching by very simple transition rules. 
The simplicity of the model allows to explore the role of the different reactions
in forming and reshaping the network and its stability with respect to both small and
extensive perturbations.
In particular, we study the convergence to equilibrium of the system as a function of the
reaction transition rates parameterized in terms of free energy differences between states.
The possibility of exploring the long-time behavior of this model in the whole parameter space
allows us to predict relaxation times ranging from seconds to hours when realistic scission free energies are considered.

Our approach to the problem of how a wormlike micellar network relaxes is crucial for pinpointing the bottleneck in the convergence. It confirms our recent suggestion~\cite{iubini2020} that the reduction of an excess of dangling ends is a difficult and hence long process. 
It involves a slow decay in which dangling ends continue to stick and detach from central parts of network strands, 
thus forming a temporary excess of branching, while eventually also annihilating in pairs from time to time. Obviously, the annihilation of dangling ends becomes an increasingly difficult process while they gradually disappear, as in standard reaction-diffusion processes with annihilation~\cite{wattis06}.
The arguments developed in this work are generic enough to be potentially relevant for characterizing colloidal systems~\cite{del05,zac06,zac08,ang14}, self-healing rubber~\cite{cor08,gre08} and network fluids~\cite{dia17}. Similar phenomena of slowing down induced by chaining effects were also observed in models of dipolar fluids~\cite{shelley99,tlusty00,miller09}.

The following section and Sec.~\ref{sec:equ} describe the model and its relaxation dynamics during a thermal quench, respectively.
In Sec.~\ref{sec:conv} we detail the properties of the relaxation to equilibrium, while Sec.~\ref{sec:eq-t} deals  specifically with equilibration times.
In Sec.~\ref{sec:mech} we show that a small ``mechanical'' perturbation leads to the same relaxation time scales of the abrupt thermal quench.
Moreover, we connect this picture to recent  experimental discoveries based on  microrheology~\cite{jai21}.
Sec.~\ref{sec:disc} is devoted to a brief summary of the main results and to some  considerations 
on the long relaxation time scales that can  occur in micellar networks.

\section{Model}
\label{sec:model}

We consider a system of interacting particles, each roughly representing the set of surfactants that would form a globular micelle.
The approach is inspired by a dynamical model of patchy particles that we have recently used for simulating the aggregation of wormlike micellar networks~\cite{iubini2020}.
However, it is important to 
realize that our mean-field model is rather general and it does not necessarily require to deal with patchy particles. Indeed, we will show that microscopic interactions between patchy particles  can be reabsorbed into effective transition rates with specific geometry-dependent prefactors that do not alter the overall relaxation features.

\begin{figure}[t!]
  \begin{tabular}{cc}
    (a)&\\
    \centering\includegraphics[width=7.0cm]{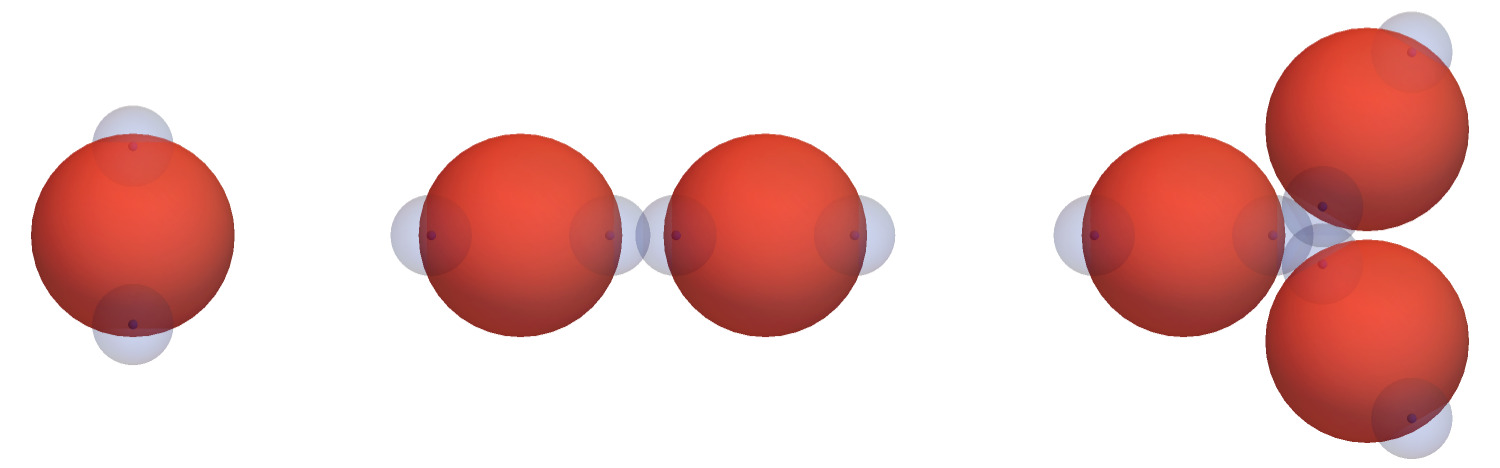} &\\
    (b)&\\
    \centering\includegraphics[width=8.4cm]{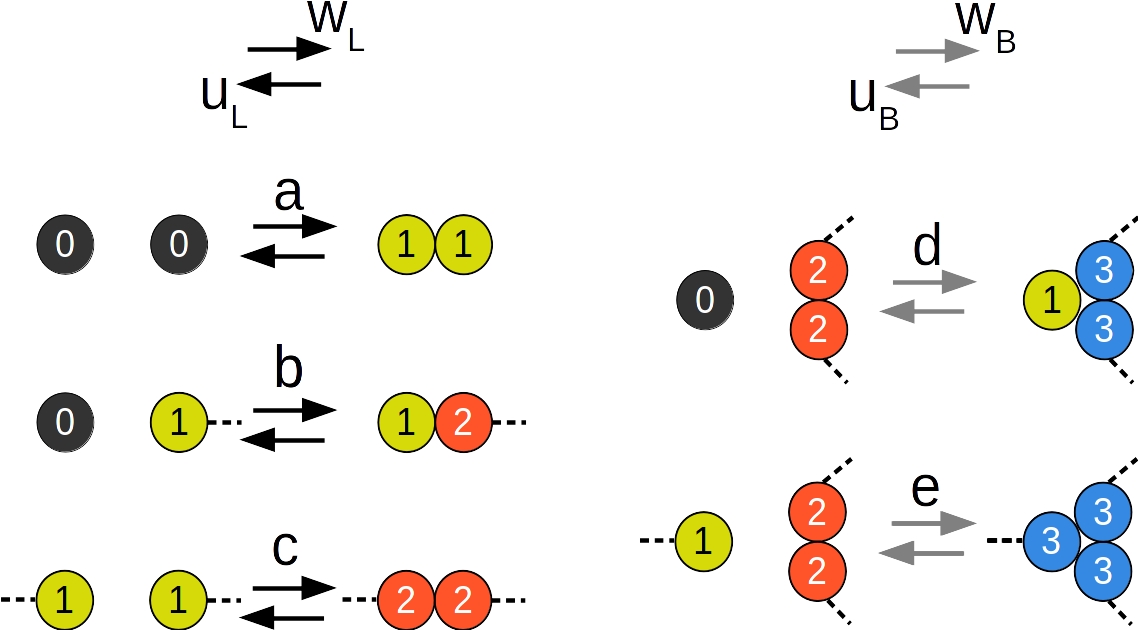}&
  \end{tabular}
  \caption{
    (a) Sketch of patchy particles, with two sticky spots (blue) and a repulsive core (red), and example of energetically favorable local contact between two particles (center) and between three particles (right).
    (b) The five transitions considered in this work, each one including synthesis (right arrows) and scission/breaking (left arrows). Each dashed line represents the continuation to a linear strand.
\label{fig:sk1}}
\end{figure}

Notice that a patchy particle is usually an engineered mesoscopic bead with sticky spots~\cite{sci09,mah16,rov18,li20}. 
Here we consider instead an idealized, nano-globule with two sticky spots at the opposite sides of its repulsive core (Fig.~\ref{fig:sk1}(a)).
Each spot is sufficiently exposed to form either one or two contacts with other particles' spots.  We focus on systems in which one contact is the typical case and two contacts per spot are the exception due to a less stable thermodynamic state.
At this scale, DNA nanostars~\cite{bif13} are another example of patchy particles, albeit with fixed valence.

At low enough temperature $T$, these particles in solution aggregate to form a  network
contributing to different parts of its motifs.
  We denote a single particle with $i$ contacts by  $\<i\>$.
  Thus, a particle $\<0\>$ is isolated, $\<2\>$ is part of a wormlike strand, $\<3\>$ is part of a branching point. Note that a particle $\<1\>$ is required, exceptionally, to have one 
free sticky spot, even if it forms two contacts with the other one, so that it always represents an end cap. 
  For a generic configuration at time $t$, we denote the number of particles $\<i\>$ by $N_{i}$
and, from the fixed total $\sum_{i}N_{i}=N$, we define the numerical fractions 
\begin{equation}
  n_{i}\equiv N_{i}/N, \qquad \vec{n}=(n_0,n_1,n_2,n_3).
  \label{eq:}
\end{equation}
Concerning the state of the system, $\vec{n}$,
we are assuming that branchings are quite rare ($n_{3}\ll 1$).
Consistently, we neglect the effects of
higher-order interactions given by more than three contacts.
Let
\begin{equation}
  p_i =\frac{n_i}{n_1 + n_2 + n_3}=\frac{n_i}{1 - n_0}
  \label{pi}
\end{equation}
be the probability that an attached particle is of type  $\<i\>$.
At sufficiently long times, when $n_0\ll n_1\ll n_3$, the normalization in~(\ref{pi}) tends to $1$, hence $p_i$ tends to $n_i$.
A set of two joined units forming, respectively, $i$ and $j$ contacts is denoted by $\<ij\>$, and of three units by $\<ijk\>$.

To derive a master equation for the evolution of the densities $\vec{n}(t)$,
we define $w_L$, $u_L$, $w_B$ and $u_B$, i.e., respectively, the rates 
for the linear synthesis, linear scission, branching synthesis and
branching scission (see the sketch in Fig.~\ref{fig:sk1}(b)).
We assume that these rates are constant and do
not depend on the local properties of the system (such as the length
of a chain), but they can depend on global quantities such as temperature
and volume fraction of surfactants.

The transitions we take into account, also sketched in Fig.~\ref{fig:sk1}(b), are
\begin{subequations}
  \label{tr}
  \begin{align}
    \<0\> + \<0\> & \xrightleftharpoons[u_L n_1 p_1/2]{w_L (2 n_0)^2/2} \<11\> \label{tra}\\
    \<0\> + \<1\> & \xrightleftharpoons[u_L n_1 p_2]{w_L (2 n_0) n_1} \<12\> \label{trb}\\
    \<1\> + \<1\> & \xrightleftharpoons[u_L n_2 p_2 /2]{w_L n_1^2/2} \<22\> \label{trc}\\
    \<0\> + \<22\> & \xrightleftharpoons[u_B n_1 p_3]{w_B (2 n_0) n_2} \<133\> \label{trd}\\
    \<1\> + \<22\> & \xrightleftharpoons[u_B n_3]{w_B n_1 n_2} \<333\> \label{tre}
  \end{align}
\end{subequations}
In the model we are neglecting transitions with very small probability to occur  (e.g.~$\<1\> + \<1\> + \<1\> \xrightleftharpoons[]{} \<333\>$) as well as
the role of unlikely motifs as the trimer $\<111\>$ shown in Fig.~\ref{fig:sk1}(a) on the right. In this respect one should regard (\ref{tr}) as an effective set of equations describing the main reaction channels of the system. 

The factor $2$ in front of each $n_0$ is related to the bivalence of each free particle, which can form contacts with either of its two sticky spots. The concentration of different kinds of particles is simply the product of their single concentrations, while there is a factor $1/2$ for each pairing between identical particles (to avoid counting twice any pairing of particles in the transitions (\ref{tra}) and (\ref{trc})).
For simplicity we are assuming that a particle $\<3\>$ is always attached to two other $\<3\>$ ones, in the rate of the backward reaction (e). At large times we will see that $n_1\ll n_3$ and the asymptotic scaling would not be affected by a more complicated scheme in which that transition rate is slightly reduced by the possibility of finding a trimer $\<133\>$ as in (d), when trying to detach a $\<3\>$ particle.

Transitions (\ref{tr}) define the corresponding fluxes
\begin{subequations}
  \label{fl}
  \begin{align}
    \Phi_a =~ & w_L 2 n_0^2   - u_L n_1 p_1 /2 \label{fla}\\
    \Phi_b =~ & w_L 2 n_0 n_1 - u_L n_1 p_2 \label{flb}\\
    \Phi_c =~ & w_L n_1^2/2   - u_L n_2 p_2 /2    \label{flc}\\
    \Phi_d =~ & 2w_B n_0 n_2   - u_B n_1 p_3 \label{fld}\\
    \Phi_e =~ & w_B n_1 n_2   - u_B n_3   \label{fle}
  \end{align}
\end{subequations}
which can be set in a column vector $\vec \Phi = (\Phi_a,\Phi_b,\Phi_c,\Phi_d,\Phi_e)^T$, so that the kinetic (nonlinear) equations for the concentrations are written in the compact form
\begin{equation}
\label{eq:kin_comp}
\frac{d {\vec n}}{dt} = {\textrm W} \vec \Phi,
\end{equation}
where
\begin{equation}
{\textrm W} = 
\begin{pmatrix}
-2 & -1 & 0 & -1 & 0 \\
2  & 0 & -2  & 1 & -1\\
0 & 1 & 2 & -2 & -2 \\
0 & 0 &  0 & 2 & 3 
\end{pmatrix}
\end{equation}
is a  $4\times 5$ stechiometric matrix
preserving the total mass (zero sum over each column).
We will see that all fluxes tend to zero during the relaxation of the system, in agreement with the expectation that thermodynamic equilibrium is established at long times.

The four transition rates $w_L$, $u_L$, $w_B$ and $u_B$ are not fully independent  but should be related by thermodynamics.
For this purpose, we introduce the dimensionless free energies $f,f_1,f_3,g$.
Terming by $\fone>0$ the energy cost of each end cap (a free sticky spot in the model),
we require that the breaking up of a linear bond, or scission, gives rise to an  increase of free energy $f = 2 \fone$.
Letting $\fthree>0$ be the cost for removing one of the three single contributions in a triple contact 
(with $\fthree<\fone$), one has that the transition from a triple contact to an exposed patch plus one linear 
contact, i.e.~the breaking of a three-hub in the network, needs an activation energy $g = 3 \fthree - 2 \fone$.
In order to have $g>0$, the constraint $\frac 2 3 \fone < \fthree $ should be imposed. 
Since in our model  a triple contact  can be at most as energetically stable as a single one, 
we also impose $\fthree\le \fone$, which leads to $g\le f/2$.
Altogether, we write 
\begin{equation}
  \label{uw}
    u_L =~  w_L~ e^{-f}, \qquad
    u_B =~  w_B~ e^{-g},
\end{equation}
and in the following we choose $(w_L, w_B, f, g)$ as the independent parameters characterizing the behavior of the kinetic equations~(\ref{eq:kin_comp}). Since  we are interested in the regime $g\ll f$ in which branching is limited,  the ``quantization'' of blobs of surfactants in particles is not crucial. As a matter of fact, in the opposite limit $g\lesssim f/2$, the  particles of the model are expected to form Kagome lattices~\cite{chen11,rom11}, 
which do not seem to be a realistic configuration for micellar networks.

Notice that the choice of writing the kinetic equations (\ref{tr}) as a function of the 
numerical fractions $n_i$ does not follow the standard approach of expressing chemical equations in terms of particle densities~\cite{gil00}.
They are proportional, in our case with identical particles, to volume fractions $\phi_i = \phi_{\rm vol} n_i$, where $\phi_{\rm vol}$ is the volume fraction of surfactants in solution.
We can, however, map  equations (\ref{tr}) to those involving the $\phi_i$'s by noticing that 
the dynamical equation of each component $n_i=n$ has the following common structure
\begin{equation}\label{eq:str}
  \frac {d}{dt} n = w~ n^2 - u~ n\,.
\end{equation}
In terms of volume fractions $\phi=\phi_{\rm vol} n$, (\ref{eq:str})  can be written as 
\begin{align}
  \frac {d}{dt} \phi = \hat w~\phi^2 - \hat u~\phi
\end{align}
with $\hat w \equiv e^{-\ln \phi_{\rm vol}} w$ and $\hat u= u$. 
For each pair of rates, $(\hat w_L,\hat u_L)$ and  $(\hat w_B,\hat u_B)$, experimental measurements of the scission free energy $\hat F$ and the branching free energy $\hat G$ fix the ratios:
\begin{align}
  \frac {\hat u_L}{\hat w_L} &= e^{-\hat F/k_B T} \\
  \frac {\hat u_B}{\hat w_B} &= e^{-\hat G/k_B T}, 
\end{align}
where the free energy costs $\hat F,\hat G$ include all enthalpic and entropic contributions of the thermodynamic processes (the addition of the entropic term better fits experimental findings~\cite{vog17}). 
In particular the entropic costs  deal with the main contribution to the mixing 
entropy coming from the free end caps or micelles, which prefer to live within 
a volume much larger than the one available when they are part
of a micellar network.

\begin{figure*}[ht]
  \includegraphics[width=8.4cm]{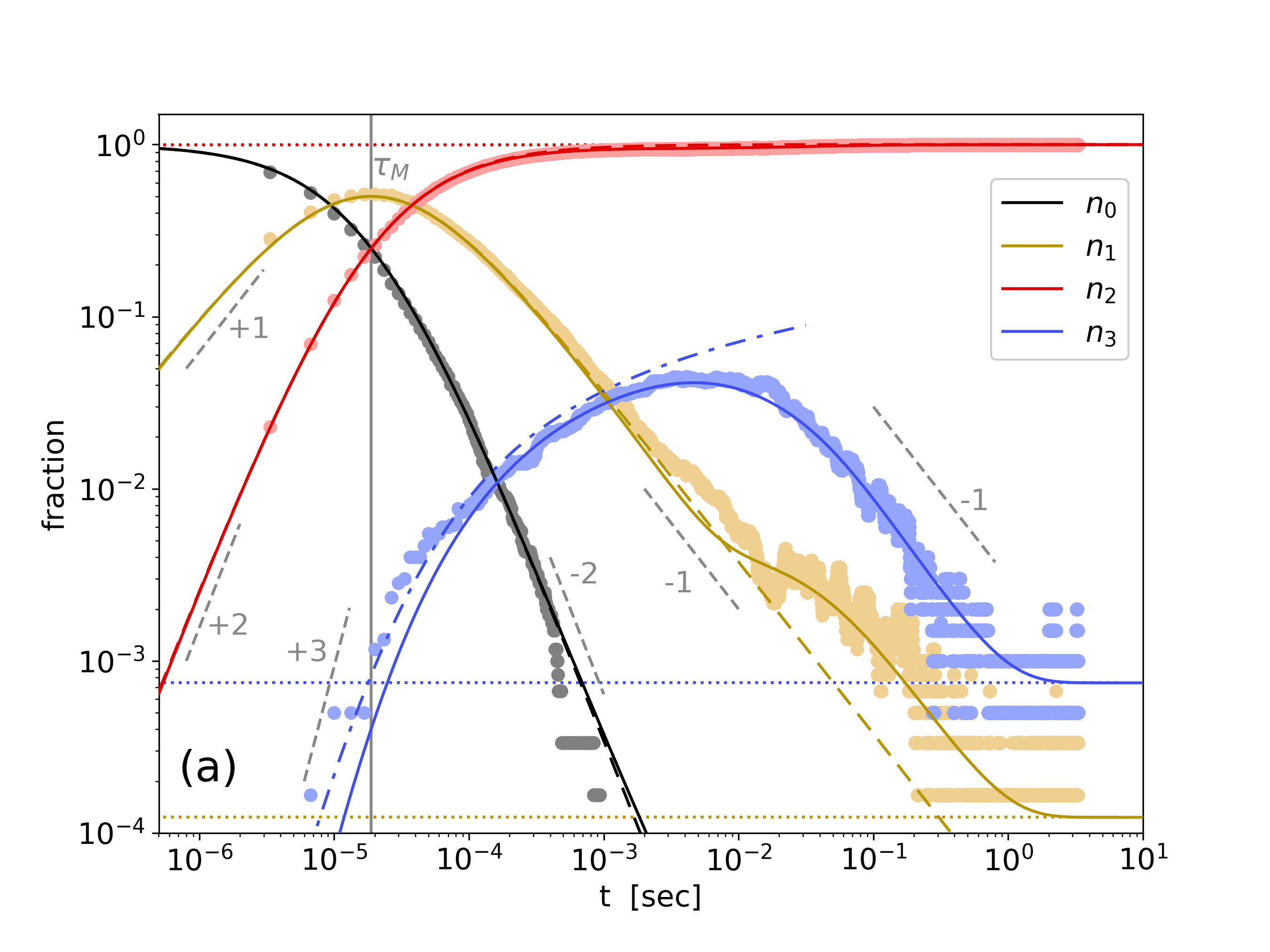}
  \includegraphics[width=8.4cm]{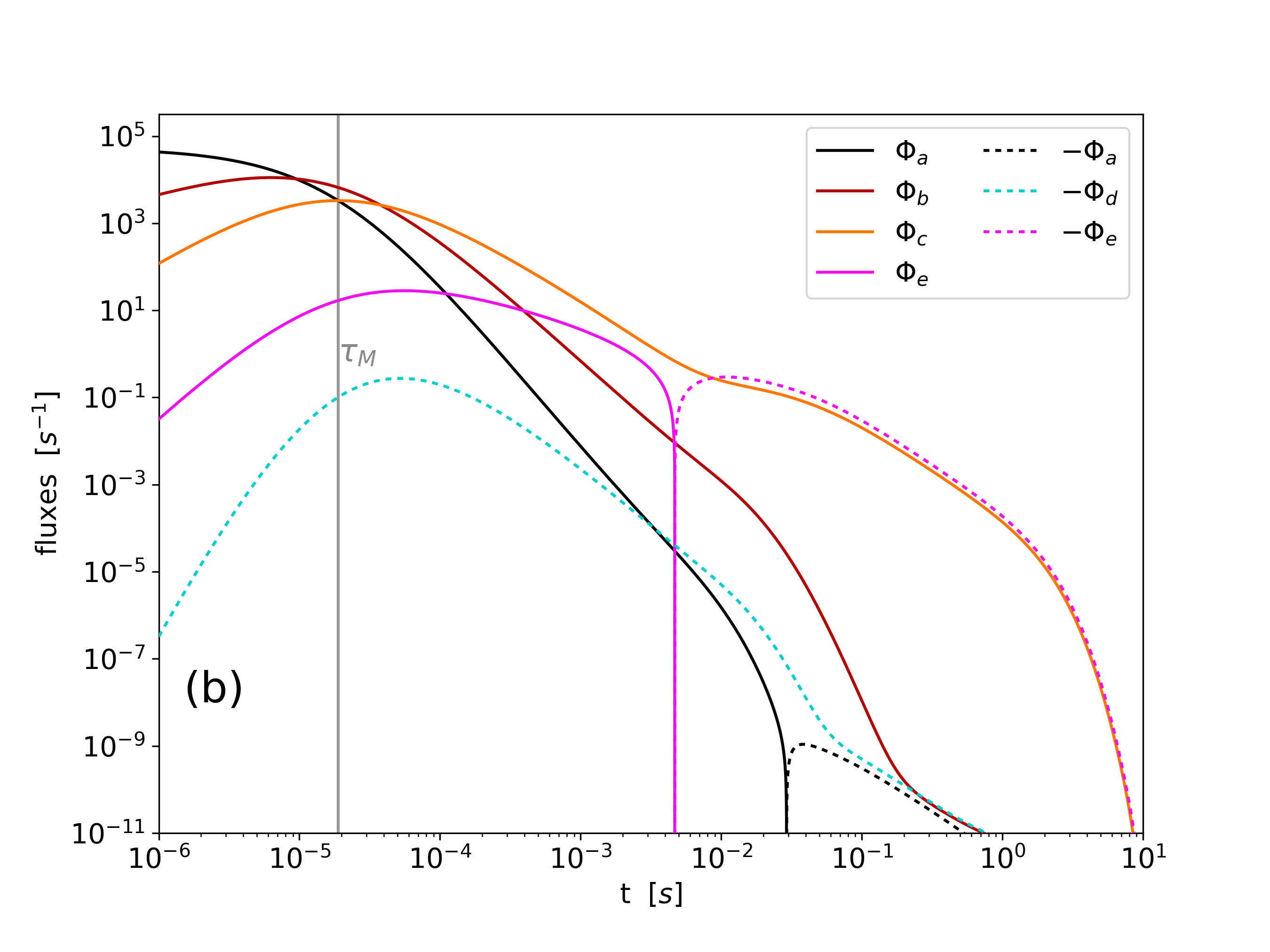}
  \caption{(a) Plot of numerical densities vs time comparing numerical data from a many-body patchy-particle model~\cite{iubini2020} (dots) with full theory integrating \eqref{eq:kin_comp} (solid curves), and with the simplified theory at short times \eqref{solsim} (dashed curves), which predicts a maximum of $n_1$ at $\tau_M=(2 w_L)^{-1} \approx 19 \mu s$ (vertical gray line). The dot-dashed line corresponds to approximating $n_3$ by \eqref{apprn3}.
      Several power-law regimes and their exponents are highlighted by gray dashed segments paired with numbers.
Parameters are $f=18$, $g=1.8$, $w_L = 26500 s^{-1}$, $w_B = 135 s^{-1}$, and simulations used $N=6000$ particles at a volume fraction $\phi_{\rm vol}=0.007$. Dotted horizontal lines correspond to approximated equilibrium values $n_1^*= e^{-f/2}$, $n_3^*= e^{g} n_1^*$ (not visible $n_0^*= e^{-f}/2$, which is too small for being detected in a system with $N\approx 10^3$ particles).
     (b) The five fluxes \eqref{fl} in log scale vs time for the same parameters of panel (a), eventually with reversed sign (see the legend). In the asymptotic decay regime we see that $-\Phi_e \approx \Phi_c$: dangling ends on average are produced from triple contact breaking and disappear by merging into linear strands. Fluxes involving $n_0$ quickly become negligible.
    \label{fig:comp}}
\end{figure*}

The relation between the parameters $f,g$ of our model and the free energy costs  $\hat F,\hat G$, whose estimates
can be obtained from experiments, reads
\begin{subequations}
  \label{convers}
  \begin{align}
    f &= \hat F/k_B T +\ln \phi_{\rm vol} \\
    g &= \hat G/k_B T +\ln \phi_{\rm vol}. 
  \end{align}
\end{subequations}
For instance, with the value  $\phi_{\rm vol}=0.007$ used in
previous numerical studies~\cite{iubini2020} (see also Fig.~\ref{fig:comp}a), 
one obtains a difference of  $\ln \phi_{\rm vol}\simeq -5$ between the experimental value
$\hat F/k_B T$ and our parameter $f$. From now on we will report the results in terms of the numerical fractions 
$n_i$, keeping in mind that \eqref{convers} may be used to convert each 
dimensionless free energy such as $f$ and $g$ to the corresponding experimental scission free energy.

The numerical integration of~(\ref{eq:kin_comp}) is performed with a standard 
fourth-order Runge-Kutta algorithm with time step $dt\le 10^{-3}$. Smaller $dt$'s are used in a range at small $t$ to achieve a constant step in log-scale.

\section{Equilibrium properties and comparison with many-body simulations }
\label{sec:equ}
\subsection{Equilibrium} 
We first determine the equilibrium values $\vec n^*$ associated to~(\ref{eq:kin_comp}).
In equilibrium, the fluxes in $\vec \Phi$ are all absent and therefore $\vec n^*$ depends only on ratios $w_L / u_L = e^f$, $w_B / u_B = e^g$.
We can perform some explicit estimates for $e^f \gg e^g > 1$, corresponding to a regime we are focusing on, where linear chains are much more stable than branches. In this situation, the statistics of $n_0$ is subordinated to that of $n_1, n_3 \ll n_2 \simeq 1$. 
Thus, we determine the approximated equilibrium solution $n_1^*, n_3^*$ by setting $n_0 = 0, n_2=1$ and keeping only the dominant terms in the conditions  $\Phi_c=0$ and $\Phi_e=0$:
\begin{align}
  \begin{cases}
    w_L n_1^2 - u_L  & = 0\\
    w_B n_1   - u_B n_3 & = 0
  \end{cases}
\end{align}
and we get
\begin{subequations}
  \begin{align}
  n_1^* & \simeq  e^{-f/2} \simeq  e^{-\fone} \label{appr:n1}\\
  n_3^* & \simeq e^{g-f/2}  \simeq  e^{3(\fthree-\fone)}.
  \end{align}
\end{subequations}
The last equation shows that $n_3$ remains a negligible fraction of the networks only if the branching free energy (per $k_B T$) $g$ has a value far lower than $f/2$. This corresponds to the case $\fthree \ll \fone$, that is when the free energy well per single contact yields a much more stable state than that given by the free energy well of a triple contact. We are not considering the scenario $g\lesssim f/2$ (i.e.~$0\ll \fthree\lesssim \fone$) in which triple contacts become abundant even if $n_1^*$ remains small.

Plugging (\ref{appr:n1}) in conditions $\Phi_a=0$, $\Phi_b=0$,  $\Phi_d=0$ yields in all cases
$n_0^* \simeq {e^{-f}} / 2$ and the approximated equilibrium solution is thus
\begin{align}
\label{eq:n*}
  n^* &\simeq \left(\frac{e^{-f}}2,\; e^{-f/2},\; 1,\; e^{g-f/2}\right)\,.
\end{align}

\subsection{Comparison with simulations}
To test the reliability of the mean field model (\ref{eq:kin_comp}) we have compared 
its time evolution   with the one obtained from 
many-body molecular dynamics simulations of a model of interacting patchy particles. 
This model is described in detail in Ref.~\cite{iubini2020}: it involves elementary rigid units composed 
of a repulsive central core of radius $R$ and two antipodal attractive spherical patches whose centers are fixed at a distance $\lambda R$.
The steric repulsion between the core beads is accounted by a shifted and truncated Lennard-Jones potential
with amplitude $\epsilon$ and spatial range $R$, while the attractive patch-patch interaction is described 
by a truncated Gaussian potential with a range equal to $0.4 R$ and amplitude $40\epsilon$. Patchy particles 
evolve in an implicit solvent at a fixed volume and constant temperature $T$.

In Fig.~\ref{fig:comp}(a) we show an example of relaxation dynamics starting from a far-from-equilibrium initial configuration of 
$N=6000$ freely diffusing isolated units $(n_0(0)=1)$ with $k_B T=\epsilon=1$ and $\lambda=1.75$ (full dots).
Solid lines represent the mean-field evolution of~(\ref{eq:kin_comp}) with parameters tuned to achieve a good fit of data ($f=18$, $g=1.8$, $w_L = 26500 s^{-1}$, $w_B = 135 s^{-1}$).
Variation of about $10\%$ of each of their values leads to worse fitting trajectories.
Remarkably, the mean-field dynamics not only relaxes  to the expected equilibrium state~(\ref{eq:n*}) in agreement with the
many-body description, but it also reproduces very well the transient
dynamics, including the non-monotonic behavior displayed by $n_1(t)$ and $n_3(t)$.
This last feature was observed~\cite{iubini2020} to produce very long relaxation time scales, especially due to
the slow dynamics of $n_3(t)$. 

Having shown that the mean field model captures extremely well 
the essential features of the micellar network dynamics, we now investigate it 
in more detail to better understand the role of the different reactions in shaping the trends of the densities, and to explore other regions of the parameters space and long 
time regimes that are practically inaccessible by simulations.

\section{Convergence to equilibrium}
\label{sec:conv}

Let us first rationalize the origin of the non-monotonic 
evolution of some fractions $n_i$ and the main features of the overall relaxation process by using simple arguments.
The maximum of $n_3$ occurring  at $t\approx 10^{-2} s$ occurs at the time at which 
the flux $\Phi_e$ changes sign (see the change from solid to dashed style of the $\Phi_e$ curve (magenta) 
in Fig.~\ref{fig:comp}(b)). 
From~(\ref{tr}), this  corresponds to the transition from 
a regime where triple contacts are favored through a dominance of forward transitions of type (e) 
to the one where its backward transition prevails.  
Since $\Phi_d$ is several orders of magnitude smaller than $\Phi_e$ we can neglect its contribution in the 
equation for ${\dot n}_3$. In this approximation
one can identify the maximum of $n_3(t)$ by imposing only the condition $\Phi_e=0$ in~(\ref{fl}),
which gives
\begin{equation}
\frac{n_3}{n_1 n_2}=e^g\,.
 \label{eq:n3max}
\end{equation}
Relation~(\ref{eq:n3max})  corresponds to a sufficient
condition for the existence  of a stationary point of $n_3(t)$ and it includes the equilibrium condition (\ref{eq:n*}) as a special case.

The dynamics of $n_1$, including its non-monotonic behavior, can be interpreted as follows.
If we neglect the contributions  of the  $\<3\>$ units  and  the 
fluxes $\Phi_d$ and $\Phi_e$, (i.e. triple contacts are forbidden),
  and we consider only the terms proportional to $w_L \gg u_L$ 
in $\Phi_a$, $\Phi_b$, $\Phi_c$, we get the simplified dynamics,
\begin{subequations}
  \begin{align}
    \frac 1{w_L} \frac d {dt} n_0 & = - 4 n_0^2 - 2 n_0 n_1\\
    \frac 1{w_L} \frac d {dt} n_1 & = 4 n_0^2 - n_1^2\\
    \frac 1{w_L} \frac d {dt} n_2 & = 2 n_0 n_1 + n_1^2
  \end{align}
    \label{dn1}
\end{subequations}
whose solution satisfying the  initial conditions $n_0(0)=1$, $n_1(0)=0$, $n_2(0)=0$ 
is
\begin{subequations}
  \begin{align}
    n_0(t) & = \frac{1}{(1+t/\tau_M)^2}\\
    n_1(t) & = \frac{2~t/\tau_M}{(1+t/\tau_M)^2}\\
    n_2(t) & = \frac{(t/\tau_M)^2}{(1+t/\tau_M)^2},
  \end{align}
  \label{solsim}
\end{subequations}
where $\tau_M =(2w_L)^{-1}$ is a typical time scale.
According to Eqs.~(\ref{solsim}), a maximum of $n_1(t)$ is reached at $t=\tau_M$ with $n_1(\tau_M)= 1/2$,  $n_0= n_2=1/4$ and $\Phi_a = \Phi_c = w_L/8$. 
Notice that these approximate results reproduce very well the early dynamics of the full system, see the position of $\tau_M$ in Fig.~\ref{fig:comp}(a,b) and the  dashed lines in Fig.~\ref{fig:comp}(a).
Furthermore, an analytical integration of $d n_3/dt$ that considers only the terms proportional to $w_B$ in $\Phi_d$, $\Phi_e$ and the solution \eqref{solsim} for $n_0,n_1,n_2$  yields
 \begin{align}
   n_3 \simeq w_B \tau_M
   \left[6 \log \left(\frac{t}{\tau_M}+1\right)
     -\frac{t}{\tau_M}\frac{29\left(\frac{t}{\tau_M}\right)^2
       +45\frac{t}{\tau_M}+18}{3 \left(\frac{t}{\tau_M}+1\right)^3}\right].
\label{apprn3}
 \end{align}
This solution neglects the constraint $\sum_i n_i=1$, yet it approximates rather well 
 the increase of $n_3$ as long as $n_3\ll 1$. Its short time expansion shows that an initial growth of $n_3\sim (t/\tau_M)^3$ ends at around $\tau_M$, see the dot-dashed line in Fig.~\ref{fig:comp}(a).

At longer times, solution \eqref{solsim} displays a power-law decay $n_1 \sim t^{-1}$,
which is typical in reaction-diffusion processes with annihilation~\cite{wattis06}.
This clarifies that the main bottleneck in the convergence to equilibrium is represented by
 the forward transition (c),
$\<1\> + \<1\> \to \<22\> $ corresponding to the merging of two dangling ends, which becomes increasingly more difficult as  $n_1$ decreases.

From Fig.~\ref{fig:comp}(a) one can also see that the presence of $\<3\>$ units interferes with the simple relaxation scaling
$n_1 \sim t^{-1}$ predicted by \eqref{solsim}.
This is indeed the regime where the approximations of \eqref{solsim} cease to be valid.
More precisely, the excess in $n_3$ at intermediate times  further
slows down the decrease of $n_1$, as the decay of triple contacts generates dangling ends through the inverse transition (e).
This is manifested by the previously discussed inversion of the flux $\Phi_e$ to negative values.
Moreover, Fig.~\ref{fig:comp}(b)  shows that, beyond the maximum of $n_3$,  
$\Phi_e \simeq -\Phi_c$, while fluxes involving $n_0$ are comparatively negligible.
Altogether, we interpret this phenomenon as the onset of a regime
in which the forward transition of type (c) (increase of strands by dangling-ends merging) is amplified by the inverse transition (e) (decrease of the number of branches).

Transition (e) takes place in both directions at time scales much faster than the decay timescale of $n_1$ ($u_B$ is not too small). Therefore, at some point $n_3$ finds its local equilibrium with $n_1$ while the latter is decreasing. This is visible in Fig.~\ref{fig:comp}(a), where $n_3$ and $n_1$ move parallel to each other, in log-log scale, toward their asymptotic equilibrium values during the last stage of the transient dynamics.
Apparently, also this asymptotic decay scales as $t^{-1}$ because the bottleneck is still the 
annihilation process $\<1\> + \<1\> \to \<22\> $ (i.e., the term proportional to $n_1^2$ in $\Phi_c$), when $n_3$ is in  thermodynamic equilibrium with $n_1$ over shorter time scales.
The picture is completed by $n_0$, whose role is marginal as it rapidly becomes much smaller than $n_1$ and $n_3$.
Correspondingly, fluxes $\Phi_a$, $\Phi_b$, and  $\Phi_d$ become much smaller in modulus than $\Phi_c, \Phi_e $.

This well defined ranking in the set of $n_i$ points at the dangling ends with $n_1$ as the main variable of the relaxation to equilibrium dynamics of the system,
with a perturbation due to the excess of triple contacts that severely slows down 
the decay of $n_1$. This is  followed by a local equilibrium between $n_3$ and $n_1$ 
during the last stage dynamics. Note that rare isolated units also remain all the time in equilibrium with the variable number of dangling ends.

\begin{figure}[t]
  \centering{}\includegraphics[width=8.4cm]{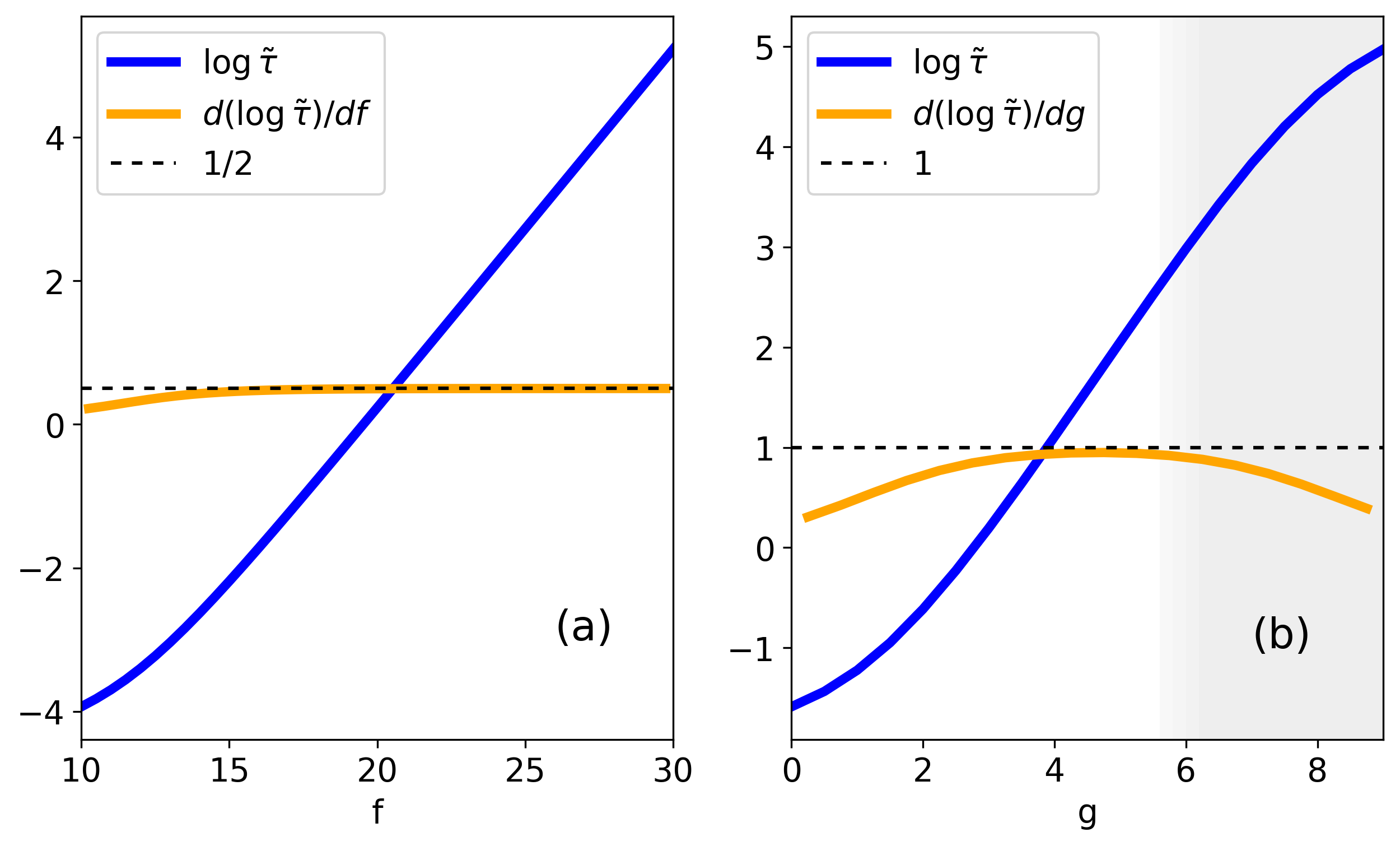}
  \caption{Log of the longest relaxation time $\tilde \tau= 1/\tilde \lambda$ obtained from the smallest eigenvalue $\tilde \lambda$ of the Jacobian matrix $J$ at equilibrium (a) as a function of $f$ with fixed $g=1.8$ and (b) as a function of $g$ with fixed $f=18$ (the shaded area indicates the region $g\lesssim f/2$ not studied in this work). In both cases we also plot the log-derivative of $\tilde \tau$, respectively vs $f$ and vs $g$. Other parameters are the same as in Fig.~\ref{fig:comp}.
    \label{fig:linsta}}
\end{figure}

\begin{figure*}[t]
  \centering{}\includegraphics[width=0.88\textwidth]{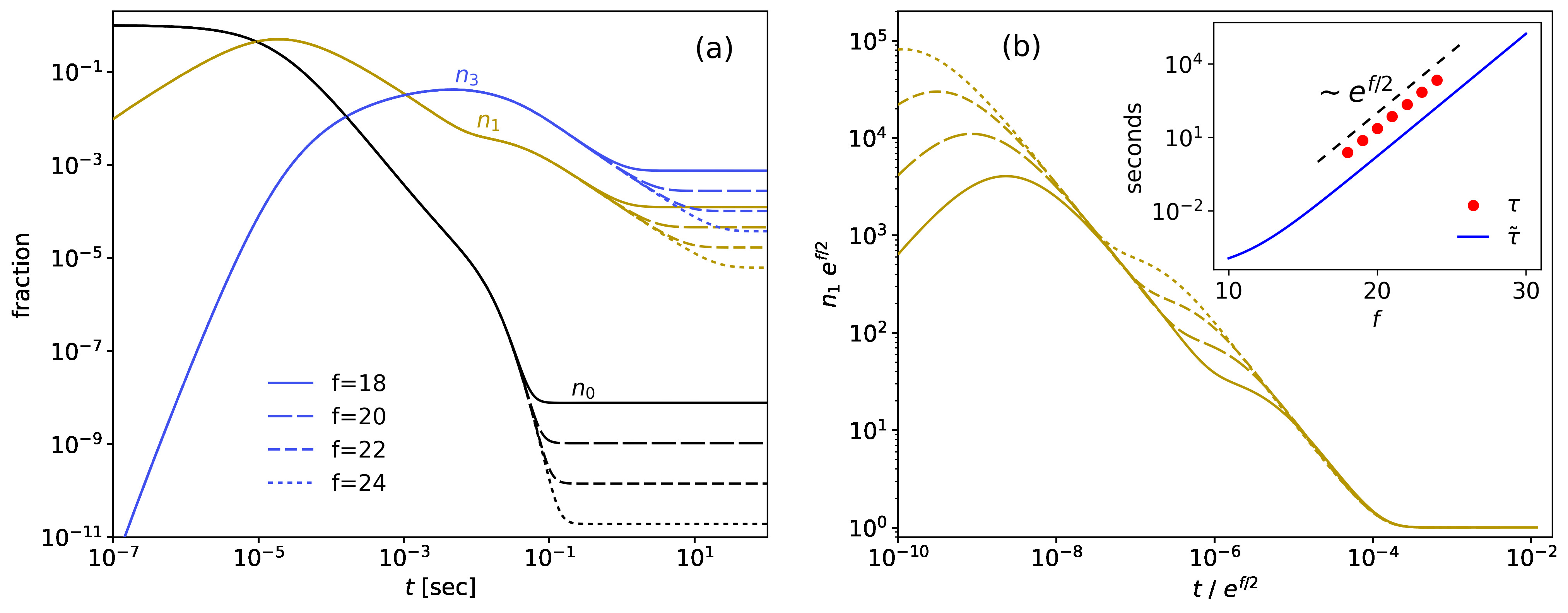}
  \caption{(a) Evolution of $n_i(t)$, $i=0,1,3$ for different values of $f$ and fixed $g=1.8$ corresponding to an initial condition $n_0(0)=1$. Other parameters are the same as in Fig.~\ref{fig:comp}, hence $\tau_M$ does not change and curves remain the same as those in Fig.~\ref{fig:comp}(a) at short times $t\lesssim 10^{-3}s$. (b)  Data collapse for $n_1$. Inset: scaling of $\tau(f)$ vs $f$ obtained from~(\ref{eq:thr}) for a threshold  $\eta =0.1$ (circles) compared with $\tilde \tau$ obtained from linear stability analysis (solid line). 
    \label{fig:f}}
  \centering{}\includegraphics[width=0.88\textwidth]{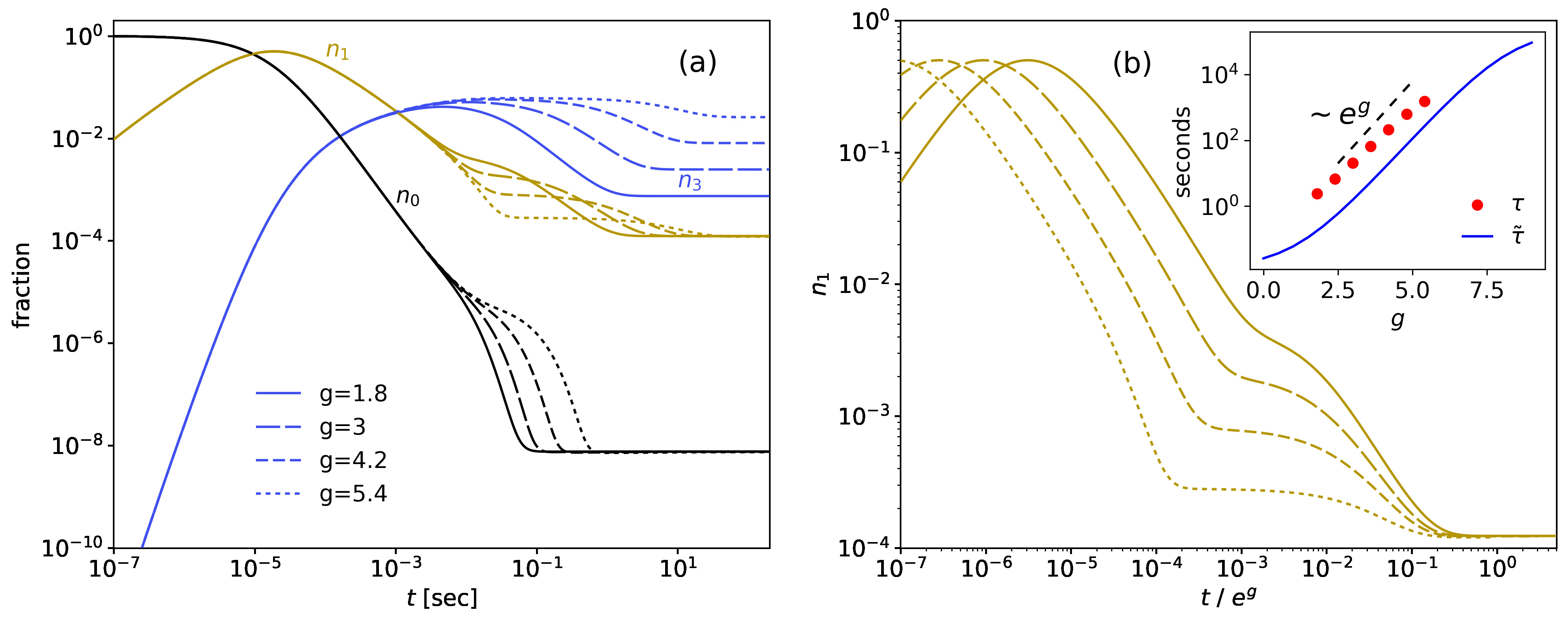}
  \caption{(a) Numerical densities vs time for several values of $g$ and fixed $f=18$ for an initial condition  
  $n_0(0)=1$. (b) $n_1$ vs time rescaled by $e^g$. Inset: scaling of $\tau$ vs $g$ obtained from~(\ref{eq:thr}) 
  for a threshold  $\eta =0.1$ (circles) compared with $\tilde \tau$ obtained from linear stability analysis (solid line).
  \label{fig:g}
  }
\end{figure*}

\section{Equilibration times}
\label{sec:eq-t}
We now explore the long-time relaxation  dynamics of the model after a thermal quench for different values of the free energies $f$ and $g$.
As a first step, we extract information from linear stability analysis around the equilibrium state $\vec n^*$.
Upon defining the small deviations $\vec{\nu}= \vec n - \vec n^*$, linearization of~(\ref{eq:kin_comp})
provides the dynamics
\begin{equation}
  \frac{d \vec{\nu}}{dt} = J \vec{\nu}\,,
  \label{nu}
\end{equation}
where $J_{i,j}=\partial ({\rm W}\Phi)_i / \partial n_j|_{\vec n=\vec n^*}$ is the Jacobian matrix  associated to~(\ref{eq:kin_comp}) evaluated at $\vec n^*$. The set $\{\lambda_i\}, i=0,\cdots,3$
of eigenvalues of $J$ provides the characteristic inverse timescales of the system. Due to the conservation of the total
mass, a vanishing eigenvalue is always present.
 Furthermore, the smallest (in absolute value) nonvanishing 
eigenvalue $\tilde\lambda$
determines the longest relaxation timescale $\tilde\tau= 1/\tilde\lambda$. In Fig.~\ref{fig:linsta} we show the behavior of
$\tilde\tau$ when $f$ and $g$ are varied independently. In both cases, we identify regions where $\tilde\tau$ scales exponentially
with $f$ and $g$, specifically $\tilde\tau\simeq \exp(f/2 +g)$. While the exponential scaling with $f/2$ is found for $f$
arbitrarily large (see its log-derivative with respect to $f$ in Fig.~\ref{fig:linsta}(a)), the scaling with $g$ is limited by the condition $g\ll f $ specified in Sec.~\ref{sec:equ}. As a result, the corresponding scaling region appears when the condition $1\lesssim g\ll f$ is satisfied, that is where $d(\log\tilde \tau) / d g \simeq 1$ in Fig.~\ref{fig:linsta}(b).

Let us now analyze the onset of slow relaxation for arbitrarily far from equilibrium states using the full nonlinear system~(\ref{eq:kin_comp}). In Fig.~\ref{fig:f}(a) we report the dynamics of $\vec n(t)$
corresponding to an initial condition where $n_0(0)=1$ and for different values of $f$ in the range $18\leq f \leq 24$
with $g=1.8$.
The evolution of $n_2(t)$ would look as in Fig.~\ref{fig:comp}(a) and has been omitted for the sake of clarity.
  We observe that the evolution of $n_i$'s is independent on $f$ until times of order $0.1 s$. Indeed, in this regime forward transitions
in Eqs.~(\ref{fl})  are dominant with respect to $f$-dependent backward transitions. On the other hand, a
suitable rescaling of the axes by a factor $e^{f/2}$ produces a data collapse in the long times region, see Fig.~\ref{fig:f}(b) for a rescaling of $n_1$. Equivalently, we can estimate the typical equilibration time $\tau$ at which $n_1$ crosses a value $10\%$ larger than the equilibrium one,
i.e.~the largest time meeting the condition 
\begin{equation}
\label{eq:thr}
n_1(\tau)/n_1^* -1= \eta\,,
\end{equation}
with $\eta=0.1$.
In the inset of Fig.~\ref{fig:f}(b) we show that $\tau(f)\sim e^{f/2}$, in agreement with the scaling of $\tilde \tau$ from the linear stability analysis.

A slightly more complex scenario emerges in Fig.~\ref{fig:g}(a) upon varying the free energy cost $g$ for breaking a hub, with fixed $f=18$.
As a first remark, the final equilibration times for moderately large (experimentally accessible) $g$'s are found to take place over minutes even for this case with small scission free energy $f$.
This final relaxation follows a stagnation period that increases with the breaking free energy $g$, in which  $n_3$ remains at a too large value compared to its equilibrium $n_3^*$.
The plateau level, however, depends mildly on $g$.
Consequently, the approach to equilibrium $n_3^*\simeq e^{g-f/2}$ from the intermediate regime is characterized by trends that scale differently by increasing $g$. This feature is clearer when time is rescaled by a factor $e^g$, as in  Fig.~\ref{fig:g}(b), where one notes that curves of $n_1(t)$ converge to the same asymptotic value $n_1^*\simeq e^{-f/2}$ but with different slopes (in log scale). Nevertheless, the scaling of $\tau\sim e^g$ obtained from~(\ref{eq:thr}) is in agreement with the results of linear stability analysis, see the comparison between $\tau$ and $\tilde \tau$ in the inset of Fig.~\ref{fig:g}(b).

By combining our findings, and by looking e.g.~at the inset of  Fig.~\ref{fig:f}(b), we see that the relaxation of the micellar network is expected to take place at a time
\begin{align}
  \label{taufg}
  \tau &\approx \tau_0 \,e^{f/2 + g}\nonumber\\
  & \approx \tau_0 \phi_{\rm vol}^2 \exp\left(\frac{\hat F}{2 k_B T} + \frac{\hat G}{k_B T}\right)
  \,.
\end{align}
The prefactor $\tau_0 \approx 40 \mu s$ can fixed by looking at Fig.~\ref{fig:comp}:
 in order to have $\tau \approx 2s$  for $f=18$ and $g=1.8$, and using \eqref{convers}, 
one gets for the  scission free energy an estimate  $\hat F \simeq 23 k_B T$ 
and  $\hat G \simeq 6.8 k_B T$ for the  branching free energy.
Clearly the stagnation in the intermediate high $n_3$ regime has a strong impact in shaping this relaxation, and rising $\hat F$ and $\hat G$ to larger values closer to experimental measurements would lead to $\tau$ values much longer than seconds. For instance, by keeping $\hat G$ fixed while raising $\hat F$ to $30 k_B T$ one gets $\tau \approx 1$ minute. If also $\hat G$ is increased  (still within the domain of our theory) say up  to $10 k_BT$, the prediction of the relaxation time gets close to half an hour.

As a curiosity, we finally mention that Fig.~\ref{fig:f}(a) is also displaying a peculiar approximated conservation law of our model: the ratio $n_3/n_1$ does not depend sensibly on $f$. Our explanation is the following: since $f$ enters in the equations only though $u_L=e^{-f} w_L$, and since $u_L$ is the smallest rate (as long as $f$ is large enough), the system's dynamics is the same for different values of $f$ till a time $\approx 0.1 s$ when a local equilibrium $n_3/n_1=e^g$ is  established. This ratio is then maintained at later times, regardless of the specific value of $f$. Therefore, the whole equilibration for different $f$'s yields the same curve for $n_3/n_1$ vs $t$ (not shown).

\section{Small mechanical perturbation and microrheology}
\label{sec:mech}

\begin{figure}[t!]
  \centering{}\includegraphics[width=8.4cm]{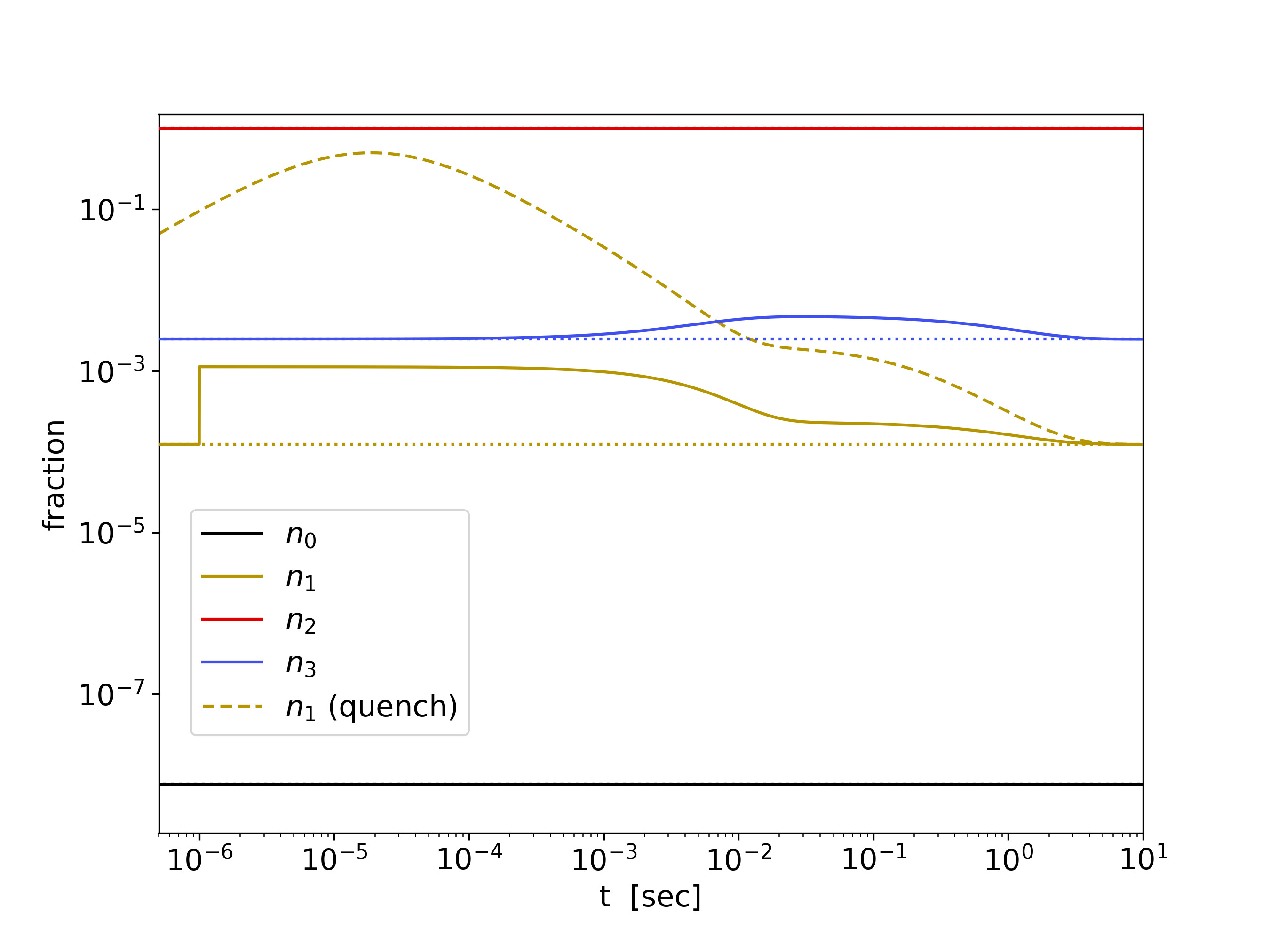}
  \caption{Relaxation of numerical densities (dense lines, see the legend) when initial conditions in equilibrium are perturbed ``mechanically'' to $n_1^*\to n_1^* + \delta$,  $n_2^*\to n_2^* - \delta$ at a small time $t_0=10^{-6}s$, for  $\delta= 10^{-3}$, $f=18$, and $g=3$.
    For comparison, we plot also the density $n_1$ (dashed line) that we obtain with the same parameters in the thermal quench. Horizontal dotted lines are equilibrium values.
\label{fig:cut}}
\end{figure}

  As one may expect, a similar relaxation scenario occurs for more general perturbation protocols.
In Fig.~\ref{fig:cut} we show the evolution arising from a gentle perturbation of particle concentrations with respect
to the equilibrium state.
In detail, we analyze the case where double contacts are slightly decreased in favor
of single contacts, i.e.~dangling ends.
We refer to this setup as a microrheological (non-thermal) perturbation, since it can be easily obtained
through a mechanical manipulation of the system whose effect is to break linear strands into shorter open chains. 
In Fig.~\ref{fig:cut}, the fraction of $n_2$ is decreased by an amount $\delta=10^{-3}$ at time $t_0=10^{-6}s$ while $n_1$ is increased
by the same amount, preserving the total density of particles (this is  the linear response regime, as similar curves are obtained with smaller $\delta$'s or by linearizing the system as in (\ref{nu})).
Overall, we observe the same global relaxation time scales as discussed in Sec.~\ref{sec:eq-t}. Interestingly, the 
mechanical setup confirms that an excess of dangling ends is again reabsorbed through a rather complex transient
dynamics that involves both the creation of isolated particles and of triple contacts, see the bumps of the black
and blue lines, respectively. Here, the slow relaxation time scale manifests itself in the rather long
duration of the bumps, which extends to the times of the order of many seconds.
This time scale matches that of a thermal quench of the kind discussed in the previous sections, as corroborated by the  comparison in  Fig.~\ref{fig:cut} with the dynamics of $n_1$ after a quench (dashed line).
All is consistent with the conclusion that even a gentle perturbation may restart the process of dangling ends generating temporarily branches (and to a lesser extent isolated micelles) and experiencing the decaying process characterized previously.

Microrheology experiments carried out at relatively slow speed of the optical tweezers display a complex oscillatory behavior~\cite{ber18,jai21} and anomalous diffusion~\cite{jeon13}. A generalized Langevin equation~\cite{dit20} can describe this behavior~\cite{ber18,jai21,doe21} if the friction memory kernel contains also an exponential term with negative prefactor. To understand this peculiar phenomenology,
it is interesting to notice that a micellar network can expose a mesh size of about $30~nm$ against dragged micro-beads of much larger diameter~\cite{ber18} $\simeq 3~\mu m$. Therefore, the forced passage of the microbead through such dense living network should create an excess of dangling ends as in the mechanical forcing ideal experiment discussed above. In fact, a clue on the peculiar motion of the bead, when dragged by optical tweezers moving at constant speed, is shown in Fig.1a of~\citet{ber18}: the observed ``intermittency'' suggests an increasing accumulation 
of local stress until the force from the traveling tweezers reaches a threshold value 
above which the living network starts to break down locally generating an excess of dangling ends.
At this point the bead starts recovering its position with respect to the center of the moving trap. It has been recently revealed~\cite{jai21} that these features  are signs of nonlinearity and, we argue, suggest a relaxation process  developing over time scales much 
longer than the structural relaxation time  $\tau_s = 2.5~s$ detected by recoil experiments~\cite{gom15}.
We also conjecture that the long time scale of the end cap recombination is difficult to measure with standard experiments.
In addition to the above phenomena, it is finally worth mentioning
 the possible role of long-range effects that were conjectured in~\cite{jeon13} to justify superdiffusive
spreading of passive tracers immersed in micellar solutions~\cite{ott90}. Understanding their role in genuinely active microrheological setups is an open interesting  problem that however requires more refined models beyond the
mean-field approximation.

\section{Discussion}
\label{sec:disc}

We have studied a mean field model for wormlike micellar networks with mild tendency to branching and we have found that the scaling of its relaxation time is modulated not only by the scission free energy of wormlike strands (as already known~\cite{tur90} from classic approaches) but also by the branching free energy.
The different route we have followed for the modeling, compared to earlier works based on the polymer length distribution, has unveiled the relevance of a perhaps so-far overlooked annihilation process, the one that gets rid of dangling ends. The possibility of merging an end cap to the interior of a wormlike strand temporarily solves its excess free energy but generates a metastable hub with three branches. There emerges an excess of these hubs over short time scales and their removal may be particularly problematic for the dynamics, especially if they are thermodynamically quite stable. As a global result, the relaxation of a wormlike micellar network may easily be very slow. The strong sensibility on the scission and braking free energies enables relaxation time scales ranging from seconds up to hours if the scission cost gets close to $30 k_B T$.

The dynamical regime we observe for the strong thermal quench involves power laws and is more structured than the simple exponential relaxations predicted in the past for small perturbations of the system. However, according to our linear response analysis, the kind of process needed to relax the system (the end-end annihilation while the branches also equilibrate) is the same either for small perturbations and large ones.

Our study confirms that the effect of branching is yet to be fully understood and that the end 
cap annihilation is a subtle long-lasting process, possibly difficult to detect and quantify in experiments. 
Assessing whether this is indeed the case would be crucial, for instance, for correctly identifying the linear 
response regime of viscoelastic fluids even in apparently gentle setups as those considered in recent microrheology experiments~\cite{jai21}.  
Finally, explicit signatures of this slow dynamics could be relevant and accessible in temperature or interfacial-tension jump  experiments~\cite{fernandez2009temporal,Reidar_et_al_PRL_2009,de2010temporal}.

\begin{acknowledgments}
We acknowledge support from Progetto di Ricerca
Dipartimentale BIRD173122/17 of the University of Padova.
Part of our simulations were performed in the CloudVeneto
platform. We thank
Martin In,
Matthias Kr\"uger,
Gianmaria Falasco,
Roberto Cerbino,
Lorenzo Boscolo Baicolo,
and Riccardo Sanson
for useful discussions.
We also thank two anonymous reviewers for their useful comments.
\end{acknowledgments}

The data that support the findings of this study are available from the corresponding author upon reasonable request.


%

\end{document}